# Floating photovoltaic systems: photovoltaic cable submersion and impacts analysis

**Ricardo Rebelo**[a,1], **Luís Fialho**[a,b,2], **Maria Helena Novais**[a,b,3]

[a]*Renewable Energies Chair, University of Évora, 7000-651 Évora, Portugal*
[b]*Institute of Earth Sciences, University of Évora, Rua Romão Ramalho, 7000-671, Évora, Portugal*
[1] ricardoar@sapo.pt
[2] lafialho@uevora.pt
[3] hnovais@uevora.pt

## Abstract

Floating photovoltaics (FPV) is an emerging technology that is gaining attention worldwide. However, little information is still available on its possible impacts in the aquatic ecosystems, as well as on the durability of its components. Therefore, this work intends to provide a contribution to this field, analysing possible obstacles that can compromise the performance of this technology, adding to an increase of its reliability and assessing possible impacts.

The problem under study is related to the potential submersion of photovoltaic cables, that can lead to a degradation of its electrical insulation capabilities and, consequently, higher energy production losses and water contamination.

In the present study, the submersion of photovoltaic cables (with two different insulation materials) in freshwater and artificial seawater was tested, in order to replicate real life conditions, when FPV systems are located in reservoirs or in the marine environment. Electrical insulation tests were carried out weekly to assess possible cable degradation, the physical-chemical characteristics of the water were also periodically monitored, complemented by analysis to detect traces of copper and microplastics in the water.

The results showed that the submersion of photovoltaic cables with rubber sheath in saltwater can lead to a cable accelerated degradation, with reduction of its electrical insulation and, consequently, copper release into the aquatic environment.

# List of Abbreviations, Acronyms, Initials and Symbols

CE-Electric conductivity (μS/cm)

FPV- Floating photovoltaic

MΩ- Megaohm (MΩ)

DO-Dissolved oxygen (% or mg/L)

ORP- Oxidation Reduction Potential (mV)

SAL-Salinity (mg/L)

TDS-Total dissolved solids (ppm)

T- Temperature (ºC)

PE- Polyethylene

PP- Polypropylene

P- Polystyrene

PMMA- Polymethyl methacrylate, plexiglass

PUR- Polyurethane

PET- Polyethylene terephthalate

PVC- Polyvinylchloride, vinyl plastics

PTFE- Polytetrafluorethylene, Teflon

# 1. Introduction

The world population is growing, causing an increase in the use of resources needed to maintain the living standard of the modern societies. The electricity consumption is increasing due to the electrification of various sectors, e.g. electric vehicles. Nearly 95% of the electricity by 2050 would need to be low-carbon, a deep transformation to achieve the global temperature change below 2°C [1]. The answer to this challenge is the use of renewable energies for the production of electricity, minimizing its ecological footprint and enabling the decarbonization of the electric system. Solar energy can provide an important share of clean electricity, either through decentralized energy production, generating energy closer to the consumption points, or with centralized power production.

The floating photovoltaic (FPV) systems allow the usage of a potentially unoccupied surface, not competing with other applications such as agriculture or urban development, particularly important factors in countries with high population density. FPV systems can also benefit from coexistence with other renewable energy sources (e.g., hydropower), taking advantage of existing infrastructures, such as electric power transmission lines, electric substations or energy storage systems (pumped-storage hydroelectricity, batteries, etc.).

In the current market there are different types of floating photovoltaic platforms from different suppliers, but in general the electrical components (modules, cables, inverters, electrical protection devices, etc.) are the same as those used in conventional PV applications on land. In general, the photovoltaic modules are installed on a plastic floating platform which makes the system buoyant. These floating systems are installed with some degrees of freedom, in order to accommodate variations in the water level and wave motion, with stability given to the platforms with a mooring and anchoring setup.

The literature about potential environmental impacts of these systems is reduced, however, some of the potential impacts on aquatic ecosystems that could arise include [2]: 1) Reduced sunlight on the reservoir / Increased heat generated – can induce changes on the water column characteristics and/or the mixing patterns of the reservoir, uneven surface heating, generate potential heat plume, reduce littoral plant/ algae growth, the biota in the limnetic zones and the primary production, increase algae decomposition rate and the oxygen demand at the bottom of the reservoir and the shading of habitats and species; 2) Reduced wind and water flow – can increase stratification and limit water mixing, reduce Dissolved Oxygen (DO) levels, depending on the reservoir covered area/ total area ratio; 3) Reduced flow in the areas surrounding the arrays – can increase the sedimentation; 4) Leaching of chemicals from the materials / Use or accidental release of oil, lubricants from boats and detergents used to clean panels – can impact the water quality and aquatic biota, and accumulate in the sediments; 5) FPV components and anchoring in the littoral and benthic zones (mooring systems, electric cables) – can destroy benthic habitats, cause direct mortality and increase the turbidity; 6) Exposure to electromagnetic fields from electric cables on the bottom and littoral zones – may have direct effects on macroinvertebrates and fish.

Therefore, it is necessary to deepen FPV systems testing and study, in order to enhance the understanding of their operation effects and optimize their use through good practices.

Furthermore, despite showing resilience to extreme phenomena of nature, some news has been published about catastrophic failures of these systems [3], [4], and FPV systems are still perceived as relatively high risk.

The first application of a floating photovoltaic system was in 2007, in Aichi, Japan, with an installed power of 20 kWp [5]. In 2008, the first commercial floating photovoltaic platform was built in a water reservoir in California, with 175 kWp [5].

After 2008, new FPV plants were installed in countries like Japan, Korea, and the United States of America. Recently, China entered this market and currently dominates the FPV sector regarding total installed capacity [5].

In Europe, this technology has had a slow growth, accelerating in the last years and currently the total installed capacity is 47.20 MW. Portugal may top this list when the construction of 50 MW in the Alentejo region is concluded, estimated to be commissioned by the end of 2021, thus increasing the total European installed capacity to 97.201 MW.

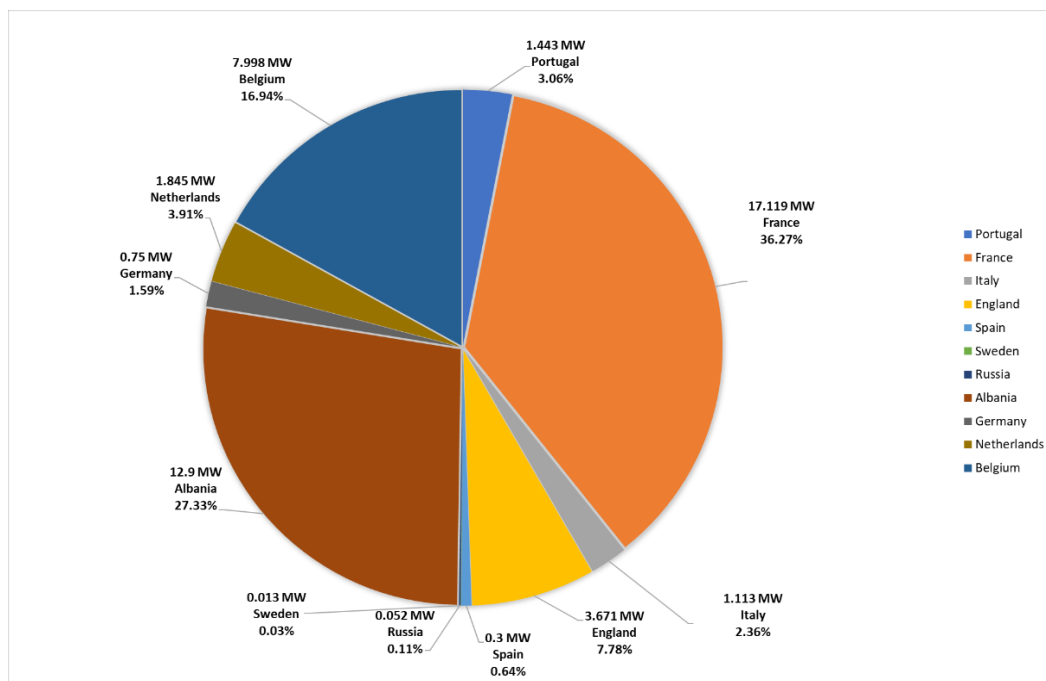

Figure 1 Floating Photovoltaic Power Installed in Europe [6], [7], [16]–[25], [8]–[15]

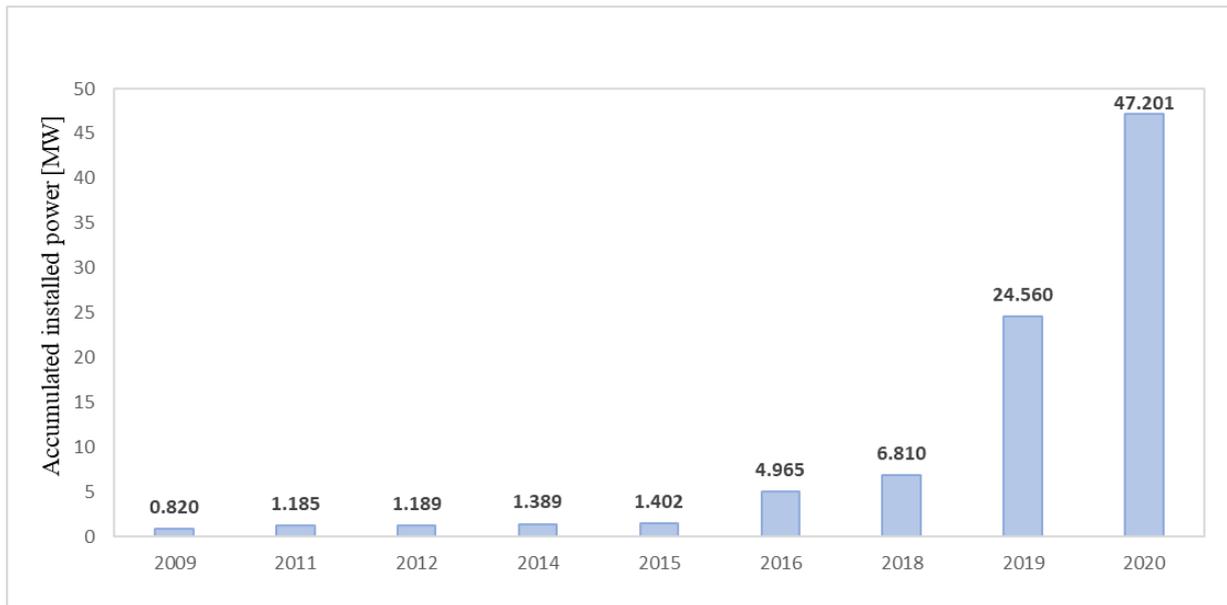

Figure 2 Evolution on FPV installed capacity in Europe [6], [7], [16]–[25], [8]–[15]

In this work, possible submersion of photovoltaic cables in water is addressed. The photovoltaic cables, that can be fully or partially submerged, will be exposed to freshwater or salt water, ice, a high humidity environment and solar radiation, which can lead to cable degradation and loss of electrical insulation. Submersion of cables or connectors can be caused by low clearance from the water surface as well as mismatch in module cable length and floats dimension, waves due to wind or a boat passing nearby.

The accelerated degradation of the photovoltaic cables or connectors can cause its failure, reducing the electric insulation characteristics or causing problems to the power line communications. In addition, individual insulation failures can often be difficult to detect in large PV systems and/or PV strings, with added difficulty related to performing O&M tasks in floating plants. In case of electric insulation failure, the photovoltaic inverters are able to detect it and will stop, isolating the faulty system. This leads to increased plant downtime, loss of energy generation and lower overall performance ratios.

The cable tests follow the EN 50618, regarding electric cables for photovoltaic systems, and EN 50395 standards, focused on electrical test methods for low voltage energy cables [26], [27].

This work intends to evaluate if the submergence of photovoltaic cables can lead to its accelerated degradation, either in freshwater or in saltwater. The degradation of the insulation layers of these cables, can lead to the direct water contact and exposure of their conductors, with potential leaching of contaminants (metals, microplastics) to the aquatic environment. This study also intends to evaluate the possibility of the occurrence of this environmental contamination.

## 2. Methodology

### 2.1. Experimental description

Currently, there are multiple types of photovoltaic cables. The conductor material is generally copper or aluminium, either solid or stranded, allowing very good conductivity, malleability, and ductility. The cable cross-sectional area and thickness of insulating layers depends on its current rating. Solar DC cables are intended for outdoor use and single-core cables with double insulation have proven to be a practical solution with high reliability in land installed PV plants.

Among the various requirements for cable selection in the photovoltaic industry, the following are often used: good weather, ozone and UV-resistance; large temperature operating range; able to withstand mechanical stress; abrasion-resistance; acid and base pH resistance; flame retardant and halogen free; high dielectric strength; small outside diameter (space-saving).

After consulting several suppliers of photovoltaic cables, two types of photovoltaic cables frequently used in photovoltaic installations (including FPV plants) in Portugal were selected, depicted in Table 1.

Table 1 Photovoltaic cables

| Cable | Conductor | Insulation and outer sheath | Section |
|---|---|---|---|
| Cable 1 | Tinned copper grade5 Compliant EN 60228 / IEC 60228 | Cross-linked polyethylene (XLPE) | 4 mm$^2$ |
| Cable 2 | Class 5 annealed and tinned electrolytic copper wires Compliant EN 60228 / IEC 60228 | Rubber | 4 mm$^2$ |

Six large tanks were used to simulate freshwater and marine environments, with a radius of 0.75m, 0.3m in height and a total volume of 0.53m3, made of high-density polyethylene POLYCHOC™ with anti UV treatment, and a food safe rating.

The six tanks were installed outdoors, filled with water, and equipped with a water mixing and oxygenation system in order to prevent algae growth and bacterial organic matter degradation. This system is composed by an air compressor and a compressed air distribution network for the six tanks. A timer system controller was installed, performing 10 daily cycles of 20 minutes, from 9:00 until 18:00. The experimental setup is presented in detail in Table 2. The outdoor distribution of the test tanks is illustrated in figure 3. This outdoor test setup was installed in the experimental campus of the University of Évora (Polo da Mitra, GPS coordinates: 38°31'53.3"N, 8°00'43.3"W). The test site is framed in a rural context, with characteristics similar to the framing of these FPV facilities (for instance the FPV 1 MWp plant in Cuba-Este reservoir, GPS coordinates: 38°10'07.2"N 7°50'51.2"W), whether climatic or regarding the surrounding ecosystem, having no major sources of atmospheric pollution or contamination in its neighborhood.

To simulate freshwater systems, tanks 1, 3 and 5 were filled with ca. 500 L of tap water, and for marine environments simulation, ca. 18kg of marine salt was added to the tanks 2, 4 and 6, to achieve a salinity level of about 3.5% (Figure 4 a). The salinity value was monitored using a TROLL 9500 PROFILER XP multi parametric probe [28] (Figure 4 b).

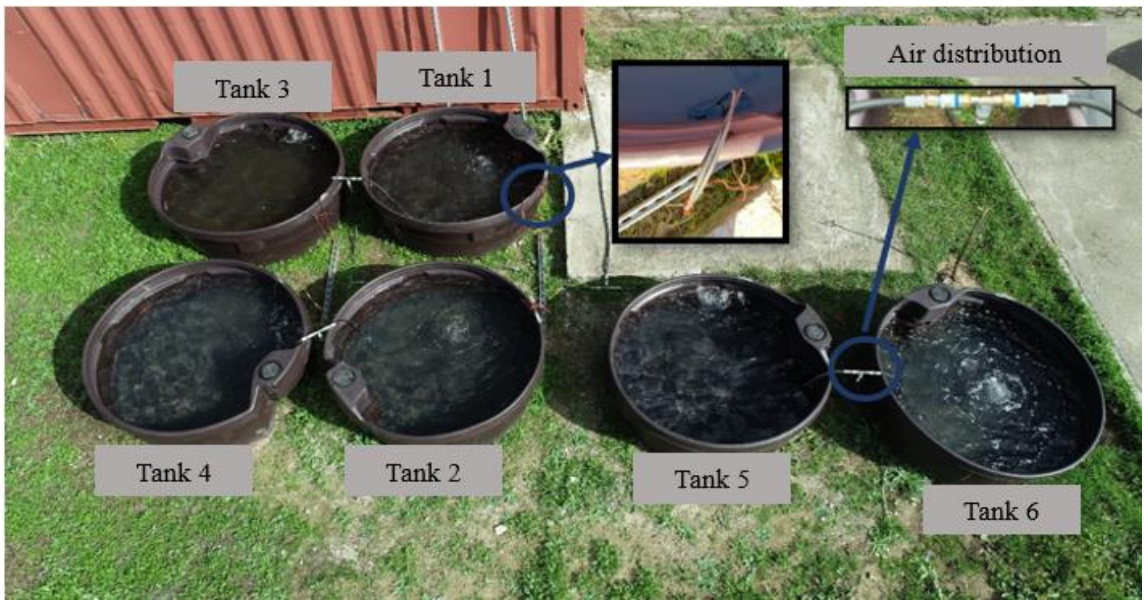

Figure 3: Experimental setup with the 6 water tanks. The air compressor and timer system are installed inside the red container (top of the image).

Table 2: Experimental setup distribution

| Tank | Photovoltaic cable | Water |
|---|---|---|
| 1 | Cable 1 | Fresh |
| 2 | Cable 1 | Salt |
| 3 | Cable 2 | Fresh |
| 4 | Cable 2 | Salt |
| 5 (control) | - | Fresh |
| 6 (control) | - | Salt |

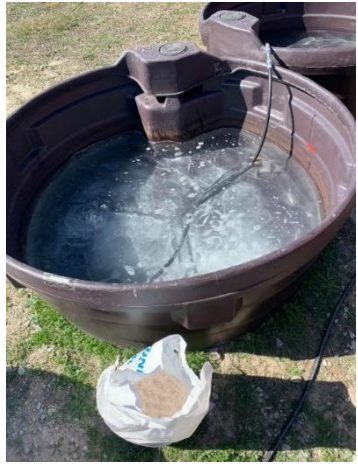 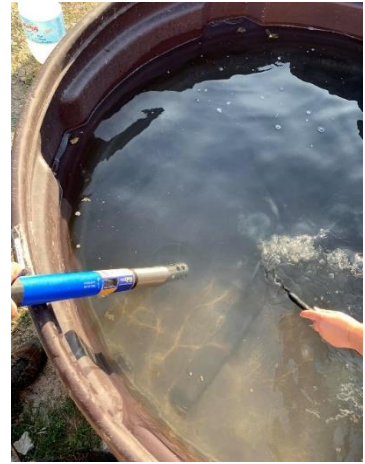

a)            b)

Figure 4 a) Preparation of saltwater; b) Physical-chemical characterization with the multiparametric probe

Weather and ambient variables were continuously monitored by a meteorological station installed 100m away from the water tanks, also in the open field. This station is equipped with a two-axis fully automatic sun tracker SOLYS2 [29], with two CMP11 pyranometers [30] to monitor global and diffuse solar radiation, one CHP1 pyreliometer [31] to measure direct solar radiation, one air temperature and relative humidity sensor [32] and one precipitation sensor [33]. This station is presented in Figure 5.

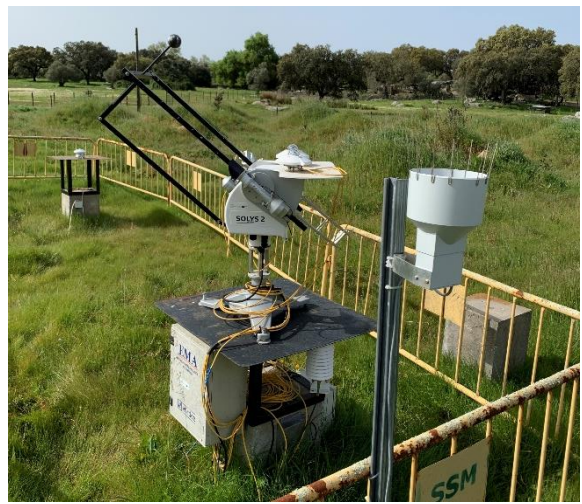

Figure 5: Onsite meteorological station.

## 2.2. Electrical insulation tests

The electrical insulation tests of photovoltaic cables follow a set of standards, namely EN 50618 and EN 50395 [26], [27]. After purchasing the cables from the supplier, and before starting the submersion test, an initial dielectric measurement was made. Each cable sample under test is 5 meters long. Special care was taken so that the ends of each cable under test were never submerged, in order to avoid the entry of water or absorption at these points, as well as the contact of the exposed conductor at the ends with water.

For the purpose of this electric tests, a voltage source [34] was used, as well as a photovoltaic and electrical installation tester [35] to measure the insulation resistance. For safety, a galvanic isolation device was also used between the electrical installer tester and the cable being measured. For measuring the cable insulation resistance, the following procedure was used:

1. Visual inspection of each cable, checking for any fault or degradation of the dielectric layer.
2. Measuring of the cable resistance with the electric tester, applying a voltage of 100 V. If the resistance result is above the minimum limit, the next step was carried out. This initial step is important in order to assess the safety conditions of the cable and allow to proceed safely in the next steps of the test.
3. As indicated by the test procedure in the standards, it is necessary to apply a voltage between 80-500 V for one minute to the photovoltaic cable submerged in water, in a small test tank, where the measurements were periodically made. For this, the voltage source was used to apply 100 V to the cable conductor and the water through a solid copper bar.
4. Finally, the cable resistance is measured with the electric tester equipment. After the resistance is measured, the cables were submerged again in the outdoor tanks. Electrical insulation tests were carried out once a week for 12 weeks.

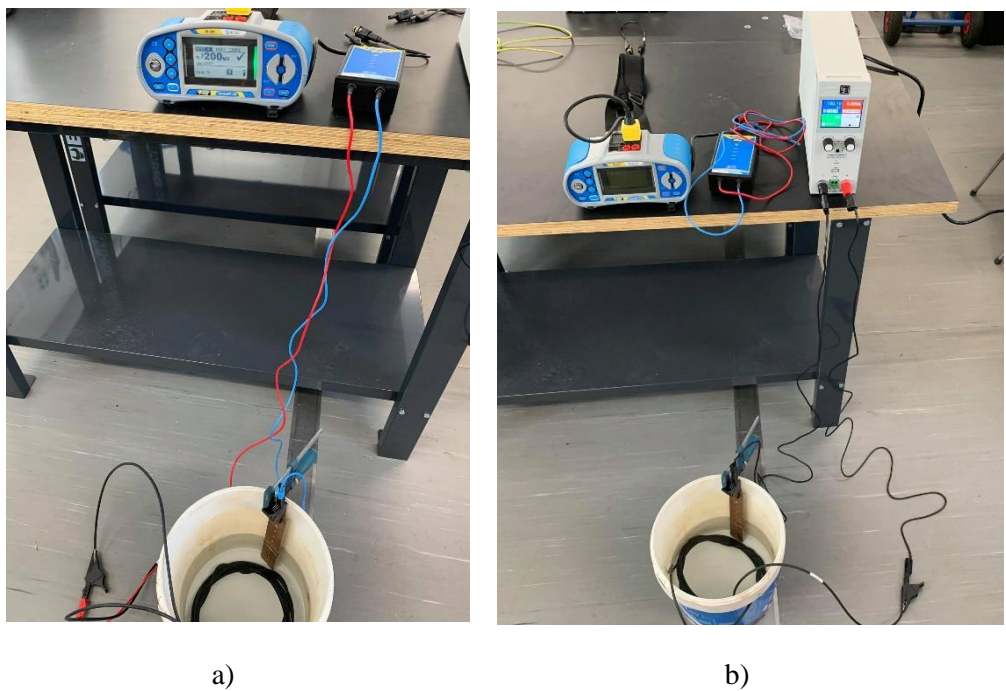

Figure 6 a) Resistance measurement with the I-V curve tracer; b) Application of 100 V with the voltage source.

## 2.3. Water physic-chemical characterization

A physical-chemical characterization of the water was carried out weekly, measuring key parameters with the multi parametric probe (TROLL 9500 PROFILER XP), including water temperature (T, °C), Electrical Conductivity (EC, µS/cm), salinity (Sal. mg $L^{-1}$), Total Dissolved Solids (TDS, mg $L^{-1}$), pH, Oxidation-Reduction Potential (ORP, mV) and Dissolved oxygen (DO, % of O2 saturation and mg $L^{-1}$). Whenever the volume of water was higher than the initial volume (due to the precipitation events), the salinity values were adjusted, removing water, and adding salt. To evaluate potential impact in the water or aquatic ecosystems due to the cable degradation, analysis to detect the presence of copper in the water through flame atomic absorption spectroscopy, using the SMEWW 3111B method [36] were carried out at the Water Laboratory of the University of Évora. This analysis was performed at different times:1) initially with the water used to fill the tanks; 2) after 6 weeks of cable submersion and 3) at the end of the test period, with 12 weeks of cable submersion.

The presence of microplastics in the water was also monitored, with the samples being filtered by a 20 µm mesh and analysed by Fourier-transform infrared spectroscopy (FTIR).

# 3. Results and discussion

## 3.1. Electrical insulation tests results

In the first measurement, all cables reached the maximum measurement limit of the device, 200 MΩ. The first cable to record electrical insulation losses was cable 2, in salt water in the eighth week, with a resistance value of 27 MΩ, and in the same week the presence of copper was detected with a significant concentration (Table 3). This decrease in the resistance value occurred when the air temperature dropped below 0°C for the first time in that year, as showed in Figure 7. The cable 2 (saltwater) presented again resistance values below 200 MΩ during the test period, nevertheless these resistance values were higher than 2.9 MΩ, the minimum insulation resistance value for a 4 mm2 cross-section cable. Cable 2 (freshwater) had also a one low resistance value, recovering to a higher value the next week. The cable 1, either submerged in freshwater, either in saltwater, never presented lower electrical resistance values.

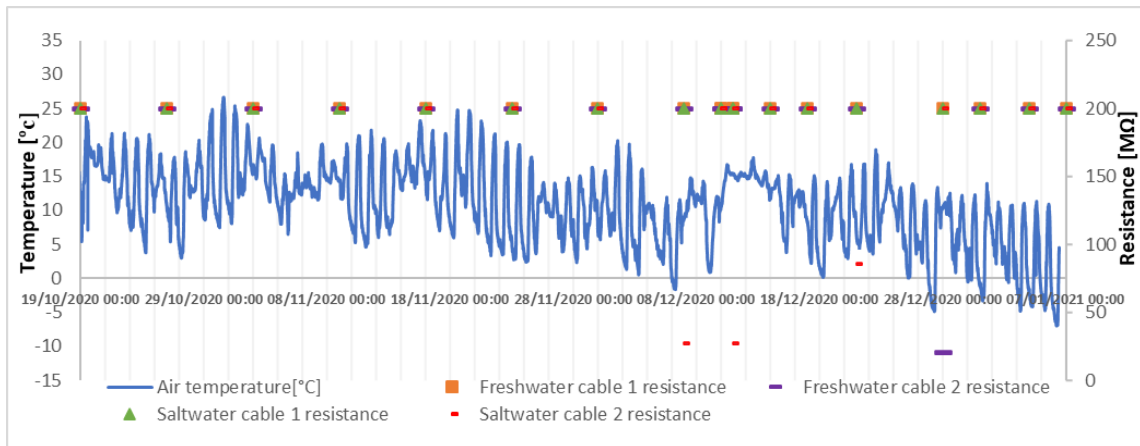

Figure 7 Air temperature and electric insulation resistance over time.

The cable 2 (saltwater) showed visual signs of degradation of the dielectric layer, with increased volume at that location, indicating potential absorption of salt water in the outer layer, as depicted in Figure 8. A possible explanation is related to the occurrence of night temperatures below 0°C and the freezing of the salt water absorbed by the outer layer, with consequent expansion of its volume and acceleration of its degradation.

That said, this points to the potential failure of this type of cable when submerged. Following this occurrence, more tests should ideally be carried out in a controlled climate chamber, to confirm this behavior.

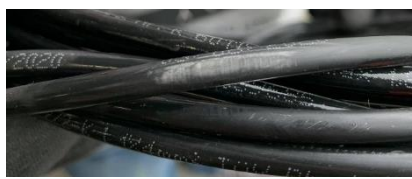

Figure 8: Cable 2 (saltwater tank) showing signs of physical degradation of the insulation layer.

## 3.2. Physical and chemical characteristics of water

The test conditions were maintained during the experimental period, in order to have aquatic environments as stable as possible, as depicted in Figure 9.

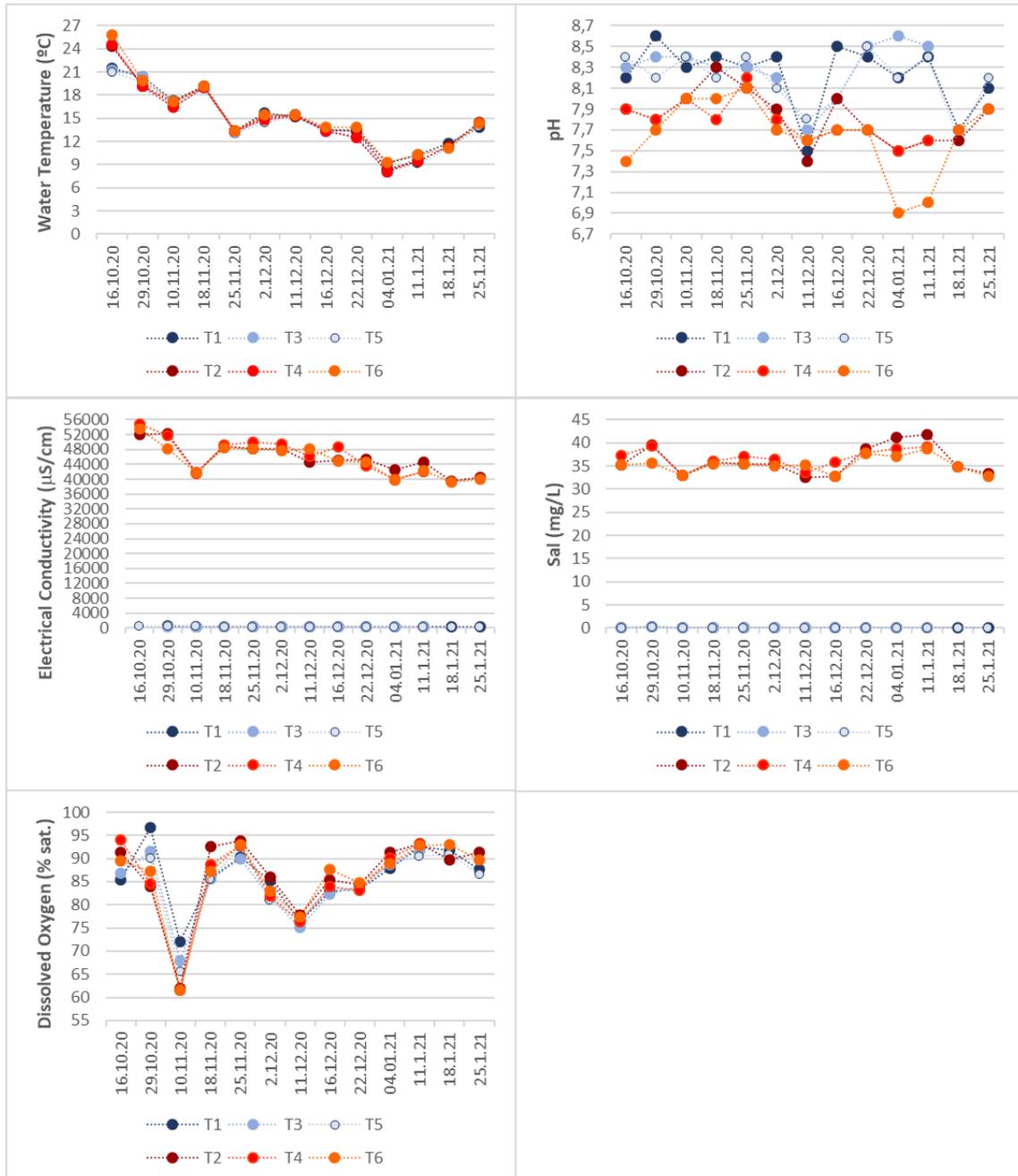

Figure 9: Physical-chemical characterization during the study period.

In Figure 9 (and in Tables S1 and S2 in Supplementary Material) is clear that most parameters were stable during the test period, and no major differences between freshwater and saltwater tanks for water temperature and dissolved oxygen were detected. Clearly, all parameters that evaluate the amount of dissolved substances and minerals in the water, as the salinity, electrical conductivity and total dissolved solids, reflect the experimental conditions, distinguishing saltwater and freshwater tanks. In spite of the efforts to keep salinity values around 35 mg L$^{-1}$, some variations can be observed (it ranges between 32.5 and 41.9 mg L$^{-1}$), since some dilution occurred due to precipitation events and sometimes an excess of salt was added since heavy precipitation events were predicted. We can also see that oxygenation levels have always been high, so there was a good circulation of water. Water temperature decreased with time, in accordance with the atmospheric temperature decrease and the seasonal changes from Autumn to Winter progression.

Regarding the possible release of copper into the water, in 3/12/2020, copper was detected in the saltwater tank 4 (cable 2) with a concentration of 0.052 mg L$^{-1}$. Since the copper concentration in the water used to fill the basins was 0.005 mg L$^{-1}$, an increase of 0.045 mg L$^{-1}$ was then registered, as can be seen in Table 3.

The decrease in copper in container 4 shortly after its detection is explained by the fact that excess water was removed from the tank due to the heavy precipitation that occurred at that time.

Table 3 Copper concentration in water over time [mg L$^{-1}$]

|  | 16/10/2020 | 03/12/2020 | 11/12/2020 | 11/01/2021 | 29/01/2021 |
|---|---|---|---|---|---|
| **Initial sample - water used to fill the tanks** | 0.005 | 0.005 | 0.005 | - | - |
| **Tank 1 - Freshwater** | 0.005 | 0.015 | - | - | 0.006 |
| **Tank 2 - Saltwater** | 0.005 | 0.008 | - | - | 0.011 |
| **Tank 3 - Freshwater** | 0.005 | 0.017 | - | 0.005 | - |
| **Tank 4 - Saltwater** | 0.005 | 0.052 | 0.012 | 0.005 | - |
| **Tank 5 – Freshwater (control)** | 0.005 | 0.014 | - | - | 0 |
| **Tank 6 – Saltwater (control)** | 0.005 | 0.009 | - | - | 0 |

In order to verify a possible water contamination with microplastic released by the cables, this analysis was carried out in all tanks. In table 4, we can observe that the microplastics concentration is always low, with the control tanks (5 and 6) presenting higher values than the others. Thus, we can conclude that there was no significant release of microplastics by the photovoltaic cables.

In future works the same analyses should be carried out, but in an aquatic environment where the remaining components of a floating photovoltaic system are, such as the floats.

Table 4 Microplastics concentration in water (The units in the table are, number of particles larger than 20 micrometres in size per litre)

| | Tank 1 | Tank 2 | Tank 3 | Tank 4 | Tank 5 | Tank 6 |
|---|---|---|---|---|---|---|
| **Organic particles (e.g., PP, PE, PS)** | - | - | - | <2 | - | - |
| **Polyethylene (PE)** | 2 | - | - | - | 2 | 2 |
| **Polypropylene (PP)** | 4 | 4 | 4 | - | 10 | 6 |
| **Organic particles (e.g., PMMA, PUR, PET)** | - | - | - | <2 | - | - |
| **Polyester** | 2 | <2 | - | - | 2 | - |
| **Ethylene-vinyl acetate (EVA)** | - | - | 2 | - | 6 | 4 |
| **Organic particles with silicone (e.g., plastic, rubber)** | <2 | <2 | <2 | <2 | <2 | <2 |
| **Organic particles with chlorine (e.g., PVC)** | <2 | <2 | <2 | <2 | <2 | <2 |
| **Organic particles with fluorine (e.g., PTFE)** | <2 | <2 | <2 | <2 | <2 | <2 |

# 4. Conclusion

The characteristics of aquatic environments pose new challenges to floating photovoltaic installations, with regard to their reliability and performance, as well as in the assessment of potential impacts on the aquatic ecosystems. The marine saltwater environment should be the most challenging for this new technology, due to its salinity, higher waves and wind speed, as well as additional anchoring and mooring difficulties. Being a recent technology, it does not yet benefit from the maturity and greater experience than conventional photovoltaic installations on land. This work intended to contribute to the increase of knowledge about these systems, in particular about the consequences of submersion of photovoltaic cables, both in the FPV installation and in the aquatic environment.

It was found that cables with rubber outer shell could fail when submerged in saltwater, with a decrease in its electrical insulation resistance and consequent release of copper into the water, potentially impacting the aquatic ecosystems. There is probably a relationship between submersion of this type of cable with temperatures below 0 ºC and the occurrence of accelerated degradation of the cable insulation layer. Reduced resistance values were measured in this cable after the occurrence of such temperatures, both in salt and freshwater. This cable type showed visible exterior degradation signs, when submerged in saltwater. However, no microplastics concentration increase was detected in the water, probably due to the cable small dimensions.

Cross-linked polyethylene insulation showed higher reliability when submerged, without presenting any reduction of its dielectric parameter. However, faults may appear for longer submersion periods, situation most likely to occur in installations in remote sites and with less frequency of inspection and maintenance visits. Further testing will be needed to determine the likelihood of failure occurring over longer submersion periods, for both cables and connectors.

The results of this work showed that it is possible the occurrence of electrical insulation failures in submerged photovoltaic cables, as well as the leaching of contaminants in case of failure. The release of microplastics has not been shown to exist with new cables, but should be revaluated for longer periods of time, with the consequent ageing of the cables.

# Acknowledgements

The authors would like to thank the support of this work, developed under the EERES4WATER project (EAPA_1058/2018), co-financed by the Interreg Atlantic Area Programme through the European Development Regional Fund. The authors also would like to thank the Water

Laboratory of the University of Évora, for the support and collaboration in the physical-chemical characterization, especially the copper analysis.

# References

[1]     "IEA, 'International Energy Agency,' 2017." https://www.iea.org/news/deep-energy-transformation-needed-by-2050-to-limit-rise-in-global-temperature (accessed Mar. 24, 2021).

[2]     P. Napier-moore, N. Cherdsanguan, L. K. Lim, and S. Merlet, "Where Sun Meets Water."

[3]     "M. Willuhn, 'The Weekend Read: Don't trow caution to the wind,' PV Magazine, 22 02 2020." https://www.pv-magazine.com/2020/02/22/the-weekend-read-dont-throw-caution-to-the-wind/.

[4]     PV Magazine, "E. Bellini, 'Japan's largest floating PV plant catches fire after Typhoon Faxai impact,'" [Online]. Available: https://www.pv-magazine.com/2019/09/09/japans-largest-floating-pv-plant-catches-fire-after-typhoon-faxai-impact/.

[5]     "Where Sun Meets Water," *Where Sun Meets Water*, 2019, doi: 10.1596/32804.

[6]     "A Study of Floating PV Module Efficiency." https://www.um.edu.mt/library/oar/bitstream/123456789/9867/1/14MSCSE012.pdf (accessed Sep. 03, 2020).

[7]     "A Review of Floating PV Installations: 2007 - 2013." https://riunet.upv.es/bitstream/handle/10251/80704/FLOAT_REVIEW.pdf?sequence=2 (accessed Sep. 03, 2020).

[8]     "Belgium's first floating PV project comes online." https://www.pv-magazine.com/2020/09/09/belgiums-first-floating-pv-project-comes-online/ (accessed Sep. 03, 2020).

[9]     "Germany Gets Its First Utility-Scale Floating Solar Power Plant!" https://www.intelligentliving.co/germany-first-utility-scale-floating-solar-power-plant/ (accessed Sep. 03, 2020).

[10]    "Ciel & Terre International – Hydrelio® floating PV system." https://www.ciel-et-terre.net/ciel-terres-technology-opens-the-doors-to-the-boot-country/ (accessed Sep. 03, 2020).

[11]    "AZALEALAAN: 1,845 KWP." https://www.ciel-et-terre.net/project/azalealaan-1845-kwp/ (accessed Sep. 03, 2020).

[12]    "Manchester to host Europe's biggest floating solar farm."

[12] https://www.businessgrowthhub.com/green-technologies-and-services/green-intelligence/resource-library/manchester-to-host-europe-s-biggest-floating-solar-farm (accessed Sep. 03, 2020).

[13] "Governantes visitam Central Fotovoltaica Flutuante de Cuba. Veja como foi este equipamento construído." https://odigital.sapo.pt/governantes-visitam-central-fotovoltaica-flutuante-de-cuba-veja-como-foi-este-equipamento-construido-c-video/ (accessed Sep. 03, 2020).

[14] "BÖR: 13 KWP." https://www.ciel-et-terre.net/project/bor-13-kwp-sweden/ (accessed Sep. 05, 2020).

[15] "Russia's first floating PV plant comes online." https://www.pv-magazine.com/2020/08/13/russias-first-floating-pv-plant-comes-online/ (accessed Sep. 03, 2020).

[16] "Albanian utility given green light for 12.9 MW floating PV plant." https://constructionreviewonline.com/news/albanian-utility-given-green-light-for-12-9-mw-floating-pv-plant/ (accessed Sep. 03, 2020).

[17] "EDP arranca com central solar flutuante no Alqueva em janeiro." https://www.dinheirovivo.pt/economia/edp-arranca-com-central-solar-flutuante-no-alqueva-em-janeiro-12774015.html (accessed Sep. 03, 2020).

[18] "FPV Alto Rabagão." https://www.edp.com/pt-pt/partilha-do-conhecimento/parque-fotovoltaico-alto-babagao (accessed Sep. 03, 2020).

[19] "A Review on Floating Solar Photovoltaic Power Plants." https://www.ijser.org/researchpaper/A-Review-on-Floating-Solar-Photovoltaic-Power-Plants.pdf (accessed Sep. 05, 2020).

[20] "O'MEGA1: FIRST FLOATING SOLAR POWER PLANT IN FRANCE." https://www.bouygues-es.com/energy/omega1-first-floating-solar-power-plant-france (accessed Sep. 03, 2020).

[21] "Viana do Castelo avalia impacto de ilhas flutuantes de energia solar no rio Lima." https://observador.pt/2020/07/08/viana-do-castelo-avalia-impacto-de-ilhas-flutuantes-de-energia-solar-no-rio-lima/ (accessed Sep. 03, 2020).

[22] "Italian engineer invents floating solar panels." https://phys.org/news/2012-02-italian-solar-panels.html (accessed Sep. 05, 2020).

[23] "SUVERETO (ITALY)." http://www.floating-solar.com/suvereto.html (accessed Sep. 03, 2020).

[24] "PIOLENC: FIRST FPV PILOT." https://www.ciel-et-terre.net/project/piolenc-first-fpv-pilot/ (accessed Sep. 03, 2020).

[25] "Floating solar panels, Godley Reservoir, UK." https://www.sciencephoto.com/media/783431/view/floating-solar-panels-godley-reservoir-uk (accessed Sep. 03, 2020).

[26] "Standards Publication Electric cables for photovoltaic systems ( BT ( DE / NOT ) 258 )," no. February, 2015.

[27] B. S. En, "Electrical test methods for electric cables —," vol. 3, 2003.

[28] "In-Situ Inc, 'Aqua TROLL 500 Multiparameter Sonde,.'" https://in-situ.com/en/aqua-troll-500-multiparameter-sonde (accessed Mar. 25, 2021).

[29] "Kipp & Zonen, 'SOLYS2 Sun Tracker,.'" https://www.kippzonen.com/Product/20/SOLYS2-Sun-Tracker#.YF9yK6_7SHt (accessed Mar. 25, 2021).

[30] "Kipp&Zonen, 'CMP11 Pyranometer,.'" https://www.kippzonen.com/Product/13/CMP11-Pyranometer#.YF9yRK_7SHt (accessed Mar. 25, 2021).

[31] "Kipp&Zonen, 'CHP1 Pyrheliometer,.'" https://www.kippzonen.com/Product/18/CHP1-Pyrheliometer#.YF9yYa_7SHt (accessed Mar. 25, 2021).

[32] "Campbell Scientific, 'EE181 Air Temperature and Relative Humidity Probe,.'" https://www.campbellsci.eu/ee181 (accessed Mar. 25, 2021).

[33] "Pronamic, 'Professional rain and precipitation sensors,.'" https://pronamic.com/ (accessed Mar. 25, 2021).

[34] "Elektro-Automatik, 'EA Programmable DC Power Supplies,.'" https://elektroautomatik.com/en/products/dc-programmable-power-supplies/ (accessed Mar. 25, 2021).

[35] "Metrel, 'MI 3109 EurotestPV Lite,.'" https://www.metrel.si/en/shop/EIS/photovoltaic-and-electrical-installation-testers/mi-3109.html (accessed Mar. 25, 2021).

[36] APHA, "American Public Health Association; American Water Works Association; Water Environment Federation," *Stand. Methods Exam. Water Wastewater*, vol. 02, pp. 1–541, 2002.

## Supplementary Material

**Table S1.** Detailed physical-chemical characterization of the tanks with freshwater during the experiment. T = water temperature; ORP = Oxidation-Reduction Potential; EC = Electrical Conductivity; Sal. = Salinity; TDS = Total Dissolved Solids; DO = Dissolved Oxygen (% saturation and mg/L)

| | Freshwater | | | | | | | | | | | | | | | | | | | | | | | |
|---|---|---|---|---|---|---|---|---|---|---|---|---|---|---|---|---|---|---|---|---|---|---|---|---|
| | T (ºC) | | | pH | | | ORP (mV) | | | EC (µS/cm) | | | Sal. (mg/L) | | | TDS (ppm) | | | DO (%) | | | DO (mg/L) | | |
| | T1 | T3 | T5 | T1 | T3 | T5 | T1 | T3 | T5 | T1 | T3 | T5 | T1 | T3 | T5 | T1 | T3 | T5 | T1 | T3 | T5 | T1 | T3 | T5 |
| **16/10/2020** | 21.5 | 21.3 | 21.0 | 8.2 | 8.3 | 8.4 | 184.3 | 180.3 | 173.3 | - | - | 602.2 | 0.0 | 0.0 | 0.0 | 0.0 | 0.0 | 0.4 | 85.5 | 86.9 | 89.2 | 7.6 | 7.8 | 8.0 |
| **29/10/2020** | 20.4 | 20.4 | 19.7 | 8.6 | 8.4 | 8.2 | 134.6 | 134.6 | 146.2 | 525.3 | 470.6 | 446.6 | 0.3 | 0.3 | 0.2 | 0.4 | 0.3 | 0.3 | 96.8 | 91.5 | 90.0 | 8.8 | 8.4 | 8.3 |
| **10/11/2020** | 17.3 | 17.2 | 16.4 | 8.3 | 8.4 | 8.4 | - | -60.0 | -64.0 | 449.0 | 469.0 | 474.0 | 0.0 | 0.0 | 0.0 | 0.2 | 0.2 | 0.2 | 72.0 | 68.0 | 65.6 | 6.3 | 5.8 | 5.7 |
| **18/11/2020** | 19.2 | 18.9 | 18.9 | 8.4 | 8.3 | 8.2 | - | -80.0 | -79.0 | 348.0 | 314.0 | 310.0 | 0.0 | 0.0 | 0.0 | 0.3 | 0.2 | 0.0 | 85.9 | 85.6 | 85.6 | 7.9 | 7.9 | 7.8 |
| **25/11/2020** | 13.3 | 13.1 | 13.2 | 8.3 | 8.3 | 8.4 | - | -61.0 | -60.0 | 420.0 | 346.0 | 285.0 | 0.0 | 0.0 | 0.0 | 0.2 | 0.2 | 0.2 | 90.4 | 90.0 | 92.5 | 9.3 | 9.3 | 9.6 |
| **02/12/2020** | 15.7 | 15.0 | 14.4 | 8.4 | 8.2 | 8.1 | - | -74.0 | -72.0 | 324.0 | 289.0 | 285.0 | 0.0 | 0.0 | 0.0 | 0.3 | 0.2 | 0.2 | 85.1 | 81.8 | 81.0 | 8.2 | 8.0 | 7.9 |
| **11/12/2020** | 15.2 | 15.4 | 15.5 | 7.5 | 7.7 | 7.8 | - | - | -37.0 | 307.0 | 278.0 | 269.0 | 0.0 | 0.0 | 0.0 | 0.2 | 0.2 | 0.2 | 76.3 | 75.1 | 76.7 | 7.3 | 7.1 | 7.2 |
| **16/12/2020** | 13.5 | 13.2 | 13.6 | 8.5 | 8.0 | 8.0 | - | - | -45.0 | 275.0 | 258.0 | 251.0 | 0.0 | 0.0 | 0.0 | 0.2 | 0.2 | 0.2 | 82.9 | 82.3 | 83.7 | 8.3 | 8.4 | 8.4 |
| **22/12/2020** | 13.5 | 12.7 | 12.4 | 8.4 | 8.5 | 8.5 | 115.7 | 106.7 | 106.2 | 250.4 | 254.0 | 240.0 | 0.0 | 0.0 | 0.0 | 0.2 | 0.2 | 0.2 | 83.4 | 83.9 | 83.1 | 9.2 | 9.5 | 9.3 |
| **04/01/2021** | 9.2 | 8.1 | 8.3 | 8.2 | 8.6 | 8.2 | 221.0 | 194.4 | 176.7 | 222.8 | 238.2 | 206.3 | 0.0 | 0.0 | 0.0 | 0.2 | 0.2 | 0.2 | 87.9 | 88.7 | 88.9 | 10.0 | 10.4 | 10.4 |
| **11/01/2021** | 10.3 | 9.9 | 9.1 | 8.4 | 8.5 | 8.4 | 168.0 | 163.8 | 174.1 | 267.0 | 243.1 | 243.2 | 0.0 | 0.0 | 0.0 | 0.2 | 0.2 | 0.2 | 92.6 | 91.8 | 90.5 | 10.4 | 10.4 | 10.4 |
| **18/01/2021** | 11.8 | - | 11.2 | 7.7 | - | 7.6 | 181.7 | - | 177.0 | 254.6 | - | 268.6 | 0.0 | - | 0.0 | 0.2 | - | 0.2 | 91.7 | - | 90.9 | 10.0 | - | 10.0 |
| **25/01/2021** | 13.8 | - | 13.9 | 8.1 | - | 8.2 | 161.3 | - | 157.1 | 221.7 | - | 210.5 | 0.0 | - | 0.0 | 0.2 | - | 0.2 | 87.6 | - | 86.6 | 9.0 | - | 8.8 |

**Table S2.** Detailed physical-chemical characterization of the tanks with salwater during the experiment. T = water temperature; ORP = Oxidation-Reduction Potential; EC = Electrical Conductivity; Sal. = Salinity; TDS = Total Dissolved Solids; DO = Dissolved Oxygen (% saturation and mg/L)

| | Saltwater | | | | | | | | | | | | | | | | | | | | | | |
|---|---|---|---|---|---|---|---|---|---|---|---|---|---|---|---|---|---|---|---|---|---|---|---|
| | T (ºC) | | | pH | | | ORP (mV) | | | EC (µS/cm) | | | Sal. (mg/L) | | | TDS (ppm) | | | DO (%) | | | DO (mg/L) | | |
| | T2 | T4 | T6 | T2 | T4 | T6 | T2 | T4 | T6 | T2 | T4 | T6 | T2 | T4 | T6 | T2 | T4 | T6 | T2 | T4 | T6 | T2 | T4 | T6 |
| **16/10/2020** | 24.3 | 24.6 | 25.8 | 7.9 | 7.9 | 7.4 | 211.6 | 201.9 | 189.3 | 52115 | 54741 | 53464 | 35.3 | 37.2 | 35.3 | 34.3 | 35.9 | 34.2 | 91.4 | 94.1 | 89.6 | 7.7 | 7.9 | 7.3 |
| **29/10/2020** | 19.2 | 19.2 | 19.9 | 7.8 | 7.8 | 7.7 | 197.0 | 190.7 | 205.3 | 52202 | 51867 | 48064 | 39.6 | 39.3 | 35.6 | 38.2 | 38.0 | 34.7 | 84.0 | 84.6 | 87.2 | 8.0 | 8.1 | 8.1 |
| **10/11/2020** | 17.1 | 16.4 | 17.2 | 8.0 | 8.0 | 8.0 | - | - | - | 41700 | 41600 | 41800 | 33.0 | 33.0 | 33.0 | - | - | - | 62.0 | 61.5 | 61.6 | 5.2 | 5.1 | 5.1 |
| **18/11/2020** | 19.1 | 19.0 | 19.2 | 8.3 | 7.8 | 8.0 | - | - | - | 48900 | 49200 | 48500 | 35.8 | 36.1 | 35.4 | - | - | - | 92.5 | 88.7 | 87.2 | 8.6 | 8.2 | 8.2 |
| **25/11/2020** | 13.4 | 13.3 | 13.3 | 8.1 | 8.2 | 8.1 | - | - | - | 48300 | 50100 | 48300 | 35.5 | 37.0 | 35.5 | - | - | - | 93.8 | 92.9 | 93.0 | 9.7 | 9.6 | 9.9 |
| **02/12/2020** | 15.2 | 14.9 | 15.5 | 7.9 | 7.8 | 7.7 | - | - | - | 48300 | 49400 | 47700 | 35.5 | 36.4 | 35.0 | - | - | - | 86.1 | 82.0 | 83.0 | 8.4 | 8.0 | 7.6 |
| **11/12/2020** | 15.4 | 15.3 | 15.4 | 7.4 | 7.6 | 7.6 | - | - | - | 44700 | 46100 | 48200 | 32.5 | 33.6 | 35.3 | - | - | - | 77.7 | 76.4 | 77.4 | 7.4 | 7.2 | 7.3 |
| **16/12/2020** | 13.5 | 13.4 | 13.8 | 8.0 | 7.7 | 7.7 | - | - | - | 45100 | 48600 | 44900 | 32.8 | 35.8 | 32.7 | - | - | - | 85.4 | 84.0 | 87.6 | 8.6 | 8.5 | 8.8 |
| **22/12/2020** | 13.3 | 12.5 | 13.9 | 7.7 | 7.7 | 7.7 | 162.2 | 158.2 | 155.5 | 45257 | 43563 | 44651 | 38.7 | 37.8 | 37.6 | 37.9 | 37.2 | 36.9 | 84.4 | 83.1 | 84.7 | 9.4 | 9.4 | 9.2 |
| **04/01/2021** | 8.4 | 8.0 | 9.3 | 7.5 | 7.5 | 6.9 | 254.0 | 243.7 | 235.8 | 42681 | 40093 | 39739 | 41.1 | 38.7 | 37.0 | 40.6 | 38.5 | 36.9 | 91.4 | 90.1 | 88.9 | 10.6 | 10.5 | 10.1 |
| **11/01/2021** | 9.4 | 9.5 | 10.3 | 7.6 | 7.6 | 7.0 | 225.2 | 220.6 | 235.8 | 44499 | 41992 | 42321 | 41.9 | 39.2 | 38.7 | 41.2 | 38.7 | 38.2 | 93.3 | 93.3 | 92.9 | 10.6 | 10.7 | 10.3 |
| **18/01/2021** | 11.4 | - | 11.1 | 7.6 | - | 7.7 | 229.7 | - | 235.8 | 39429 | - | 39278 | 34.8 | - | 34.9 | 34.6 | - | 34.7 | 89.8 | - | 93.1 | 10.0 | - | 10.5 |
| **25/01/2021** | 14.5 | - | 14.4 | 7.9 | - | 7.9 | 237.7 | - | 222.5 | 40593 | - | 40020 | 33.3 | - | 32.8 | 33.0 | - | 32.6 | 91.4 | - | 89.8 | 33.3 | - | 32.8 |